\documentclass[]{aa} % normal version
\usepackage{amsmath}
\usepackage{graphicx}
\usepackage{graphics}
\usepackage{subfigure}
\usepackage{txfonts}
\usepackage{natbib}
\bibpunct{(}{)}{;}{a}{}{,}

\def\teff{\ifmmode T_{\rm eff} \else $T_{\mathrm{eff}}$\fi}

\def\ltsima{$\buildrel<\over\sim$}
\def\lsim{\lower.5ex\hbox{\ltsima}}
\newcommand{\hi}{H~{\sc i}}

\newcommand{\ha}{\ifmmode {\rm H}\alpha \else H$\alpha$\fi}
\newcommand{\hb}{\ifmmode {\rm H}\beta \else H$\beta$\fi}
\newcommand{\lya}{\ifmmode {\rm Ly}\alpha \else Ly$\alpha$\fi}

\newcommand{\ebv}{\ifmmode E_{\rm B-V} \else $E_{\rm B-V}$ \fi}
\def\micron{$\mu$m}

\def\msun{\ifmmode M_{\odot} \else M$_{\odot}$\fi}
\def\msunyr{\ifmmode M_{\odot} {\rm yr}^{-1} \else M$_{\odot}$ yr$^{-1}$\fi}
\def\zsun{\ifmmode Z_{\odot} \else Z$_{\odot}$\fi}
\def\lsun{\ifmmode L_{\odot} \else L$_{\odot}$\fi}

\def\mup{\ifmmode M_{\rm up} \else M$_{\rm up}$\fi}
\def\mlow{\ifmmode M_{\rm low} \else M$_{\rm low}$\fi}
\def\bacs{B$_{\rm 435}$}
\def\vacs{V$_{\rm 606}$}
\def\iacs{i$_{\rm 776}$}
\def\zacs{z$_{\rm 850LP}$}

\newcommand{\oh}{\ifmmode 12 + \log({\rm O/H}) \else$12 + \log({\rm
O/H})$\fi}
\def\hyperz{{\em Hyperz}}
\def\flyf{\ifmmode f_{\rm Lyf} \else $f_{\rm Lyf}$\fi}
\def\pz{\ifmmode P(z) \else $P(z)$\fi}
\def\ki2{\ifmmode \chi^2 \else $\chi^2$\fi}
\def\zphot{\ifmmode z_{\rm phot} \else $z_{\rm phot}$\fi}

\newcommand{\xphot}{\ifmmode x_\gamma \else $v_\gamma$\fi}
\newcommand{\xobs}{\ifmmode x_{\rm obs} \else $x_{\rm obs}$\fi}
\newcommand{\xcmf}{\ifmmode x_{\rm CMF} \else $x_{\rm CMF}$\fi}
\newcommand{\vexp}{\ifmmode V_{\rm exp} \else $V_{\rm exp}$\fi}
\newcommand{\vmax}{\ifmmode V_{\rm max} \else $V_{\rm max}$\fi}
\newcommand{\nh}{\ifmmode N_{\rm HI} \else $N_{\rm HI}$\fi}
\newcommand{\dv}{\ifmmode \Delta v({\rm em-abs}) \else $\Delta v({\rm em}-{\rm abs})$\fi}

\def\fesc{\ifmmode f_{\rm esc} \else $f_{\rm esc}$\fi}
\def\flya{$f_{\rm{Ly}\alpha}$}
\def\frellya{\ifmmode f^{\rm rel}_{\rm{Ly}\alpha} \else $f^{\rm rel}_{\rm{Ly}\alpha}$\fi}

\def\ewlya{$EW({\rm{Ly}\alpha})$}

\def\hi{H{\sc i}}

\begin{document}
  \title{On \lya\ emission in $z \sim$ 3--6 UV-selected galaxies}
  \subtitle{}
  \author{Daniel Schaerer\inst{1,2}, Stephane de Barros\inst{1}, Daniel P. Stark\inst{3}}
%  \offprints{}
  \institute{
Observatoire de Gen\`eve, Universit\'e de Gen\`eve, 51 Ch. des Maillettes, 1290 Versoix, Switzerland
         \and
CNRS, IRAP, 14 Avenue E. Belin, 31400 Toulouse, France
%Laboratoire d'Astrophysique de Toulouse-Tarbes, Universit\'e de Toulouse, CNRS, 14 Avenue E. Belin, 31400 Toulouse, France
\and 
Kavli Institute of Cosmology and Institute of Astronomy, University of Cambridge, Madingley Road, Cambridge CB30HA, UK
 }

\authorrunning{}
\titlerunning{On \lya\ emission in $z \sim$ 3--6 UV-selected galaxies}

\date{Received date; accepted date}

%\abstract{%CONTEXT}
{%AIMS
%}
%{%METHODS
%}
%{%RESULTS
%}
%{%CONCLUSIONS
%}
% 5 {} token are mandatory
\abstract{Determining \lya\ properties of distant galaxies is of great interest for various astrophysical 
studies.}
{We examine how the strength of \lya\ emission can be constrained from broad-band SED fits instead 
of relying on spectroscopy.}
{We use our SED fitting tool including the effects of nebular emission, considering in 
particular \lya\ emission as a free parameter,  and we demonstrate our method with simulations of 
mock galaxies. Using this tool we analyse a large sample of U, B, V, and i dropout
galaxies with multi-band photometry.}
{We find significant trends of the fraction of galaxies with \lya\ emission
increasing both with redshift $z$ and towards fainter magnitude (at fixed $z$), and similar trends for the
\lya\ equivalent width. 
Our inferred \lya\ properties are in good agreement with the available spectroscopic observations
and other data.}
{ These results demonstrate that the strength of \lya\ emission in distant star-forming galaxies can be inferred quantitatively from broad-band SED fits, at least statistically for sufficiently large samples with a good photometric coverage.}

  \maketitle

%%%%%%%%%%%%%%%%%%%%%%%%%%%%%%%%%%%%%%%%%%%%%%%%%%%%%%%%%%%%%%%%%%%%%%%%%%%%%%%%%
\section{Introduction}
For various reasons observations of the \lya\ line in high redshift galaxies are of great interest.
For example, \lya\ together with other observations can provide useful information on the
physical properties of distant galaxies, such as their outflows, dust content, \hi\ column density,
age and others \citep[e.g.][]{shapley03,verhamme08,atek08,pentericci09,hayes10b,hayes11}
It may also be used as a signature to distinguish ``normal'' stellar populations from
extremely metal-poor ones, or even to find Population III stars \citep[e.g.][]{schaerer03,
nagao08}.
Understanding the behaviour of \lya\ also allows us to clarify the overlap  between different
galaxy types such as Lyman break galaxies (LBGs) and Lyman alpha emitters (LAEs)
\citep{verhamme08} and the influence of this line on selection functions
\citep{stanway08,reddy09}.
Finally, \lya\ observations of distant galaxies are also being used to study questions
of more cosmological nature, such as galaxy clustering and dark energy, and to
probe cosmic reionisation beyond $z>6$ \citep{hamana04,malhotra04,hill08}.

Generally,  the \lya\ properties of distant galaxies have been determined from spectroscopic
observations (follow-up or blind) or via narrow-band observations targetted to specific
redshifts. 
\citet{cooke09}  has shown that LBGs at $z \sim 3$ dominated by \lya\ absorption or emission 
can be distinguished based on broad-band photometry. 
Using SED modeling techniques taking emission lines into account
\citep[cf.][]{schaerer&debarros2009,schaerer&debarros2010}, we take this one step further 
demonstrating quantitatively that the properties of \lya\ emission can also be inferred from 
broad-band observations,  
at least statistically for large samples with sufficient photometric bands.
This allows us, for example, to determine trends
of \lya\ with redshift and other parameters, without resort to spectroscopy.

Our paper is structured as follows. The observational data and the method used for SED modeling
are described in Sect.\ \ref{s_models}. The results are presented and discussed in Sect.\ \ref{s_results}.
Sect.\ \ref{s_conclude} summarises our main conclusions.
We adopt a $\Lambda$-CDM cosmological model with $H_{0}$=70 km s$^{-1}$ Mpc$^{-1}$, 
$\Omega_{m}$=0.3 and $\Omega_{\Lambda}$=0.7. 

%%%%%%%%%%%%%%%%%%%%%%%%%%%%%%%%%%%%%%%%%%%%%%%%%%%%%%%%%%%%%%%%%%%%%%%%%%%%%%%%%
\section{Observational data and SED modeling}
\label{s_models}

\citet{dBetal11} have analysed in depth a large sample of $z \sim$ 3--6 dropout
selected galaxies using an up-to-date photometric redshift and SED fitting tool,  which treats the 
effects of nebular emission on the SEDs of galaxies. 
In their homogeneous analysis they determine the main physical properties, such
as the star formation rate (SFR), stellar mass, age, and reddening.
They assess carefully their uncertainties, and discuss the evolution of these properties with redshift.
We here extend these simulations to perform SED fits with variable strengths of 
\lya\ for the same set of galaxies, with the aim of examining whether
the available photometry allows us to distinguish any trends of \lya\ with
redshift and/or other properties.

\subsection{Photometric data and sample selection}
We have used the GOODS-MUSIC catalogue of \citet{santini09} 
providing photometry in the U, \bacs, \vacs, \iacs, \zacs, J, H, K, bands
mostly from the VLT and HST, and the 3.6, 4.5, 5.8, and 8.0 \micron\ bands
from the IRAC camera onboard {\em Spitzer}.
Using standard criteria as in \citet{stark09} we have then selected U, B, V, and i-drop 
galaxies. To reduce the contamination rate (typically $\sim$ 10--20 \%) we
have only retained the objects whose median photometric redshifts agree 
with the targetted redshift range. This leaves us with a sample of 389, 705, 199, and 60 
galaxies with median photometric redshifts of $\zphot= 3.3$, 3.9, 4.9, and 6.0.
See \citet{dBetal11} for more details.

% % % % % % % % % % % % % % % % % % % % % % % % % % % % % % % %
\begin{figure*}[htb]
\centering
\includegraphics[width=8.8cm]{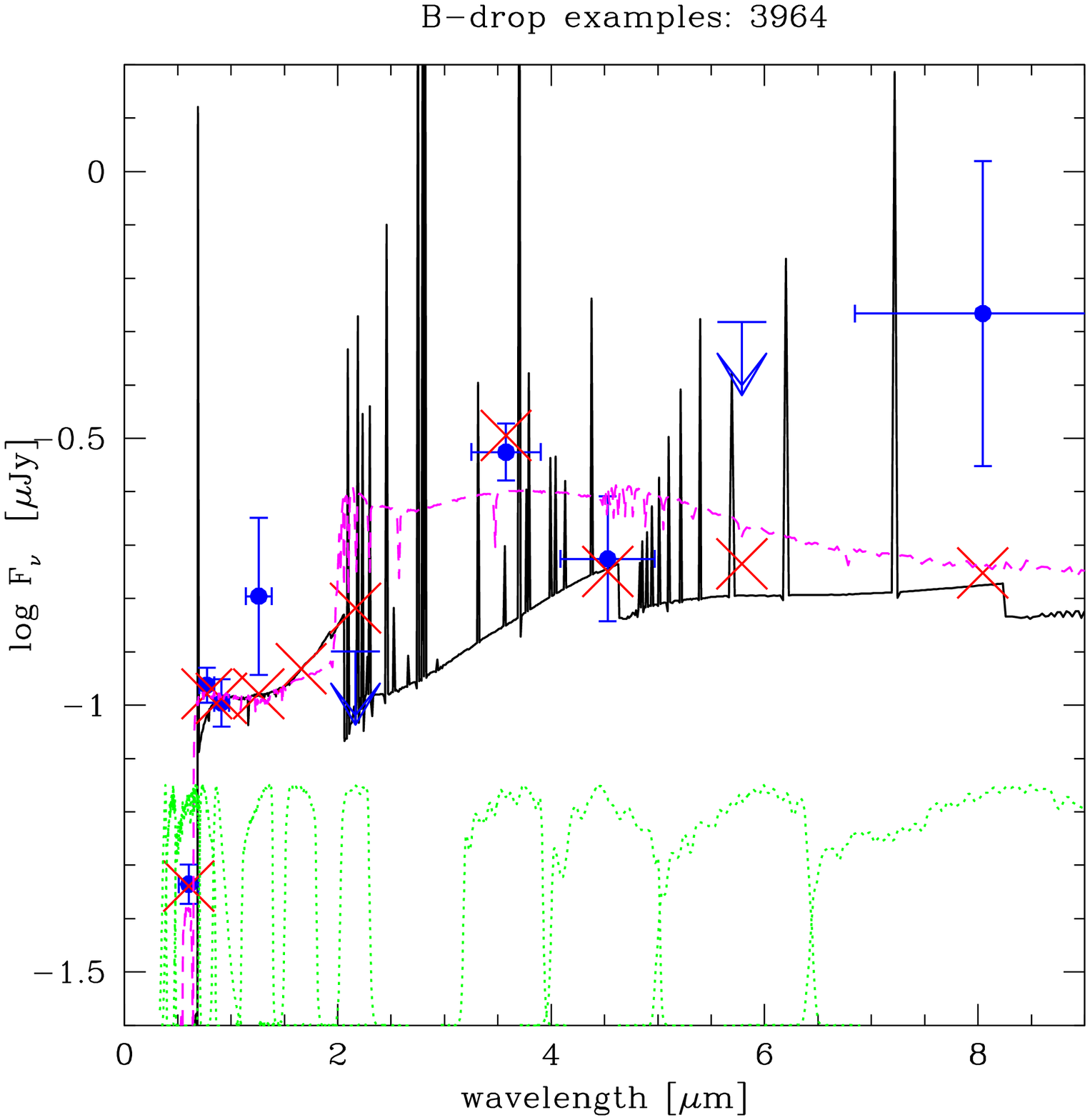}
\includegraphics[width=8.8cm]{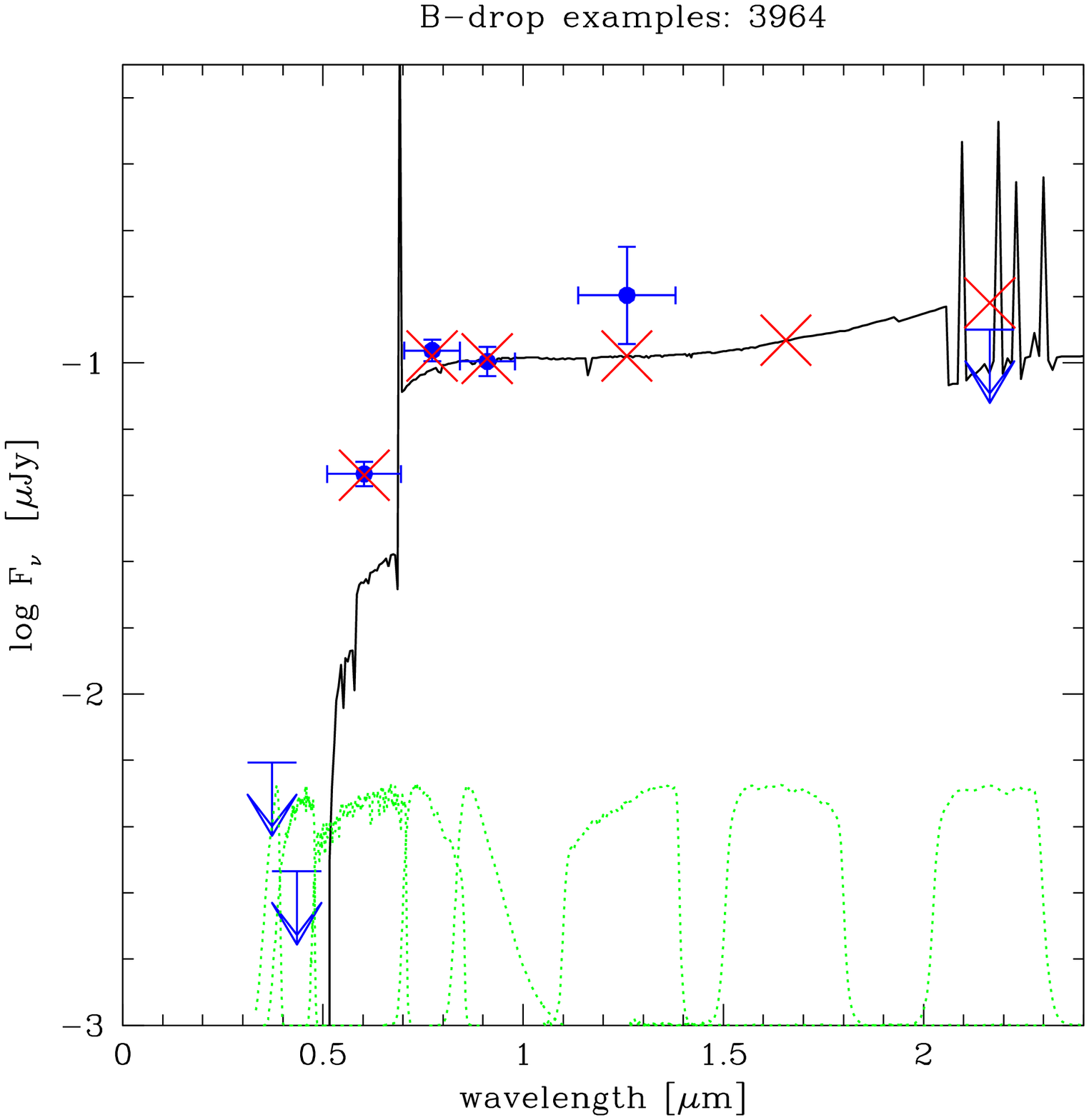}
\caption{Example of an SED fit for B-drop galaxy best fit with strong \lya.
Black solid line: best-fit SED including nebular emission.
Red crosses: flux in the photometric bands derived from the SED including
nebular emission.
Blue symbols show the observed fluxes or 1 $\sigma$ upper limits.
Green dashed lines show the positions/shape of all the availabe photometric
bands.
{\bf Left:} Full SED showing all the available photometric data.
{\bf Right:} Zoom on the SED showing the spectral range up to 2.4 \micron\
only (restframe UV to Balmer break).
 }
\label{fig_sed1}
\end{figure*}

\begin{figure}[htb]
\centering
\includegraphics[width=8.8cm]{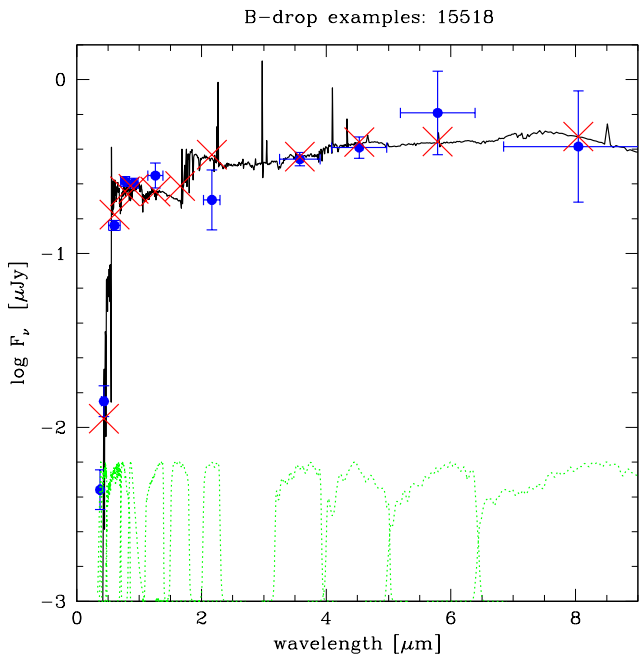}
\caption{Same as Fig.\ \protect\ref{fig_sed1} for an object best-fit with a low \ewlya.
 }
\label{fig_sed2}
\end{figure}
% % % % % % % % % % % % % % % % % % % % % % % % % % % % % % % %

\subsection{SED models}
Our SED fitting tool, described in \cite{schaerer&debarros2009} and \cite{schaerer&debarros2010},
is based on a version of the \hyperz\ photometric redshift code of \cite{bolzonellaetal2000}, 
modified to taking into account nebular emission.
In \citet{dBetal11} we consider a large set of spectral templates based on
the GALAXEV synthesis models of \cite{bruzual&charlot2003}, covering
different metallicities and a wide range of star formation
histories. A Salpeter IMF is adopted.
Nebular emission from continuum processes and numerous emission lines is added to the
spectra predicted from the GALAXEV models as described in
\cite{schaerer&debarros2009}, proportionally to the Lyman continuum
photon production. 
The intergalactic medium (IGM) is treated with the prescription of \citet{Madau95}.

The free parameters of our SED fits are:
% \citep[see][]{dBetal11}: 
redshift $z$,  metallicity $Z$ (of stars and gas),  star
formation history described by the timescale $\tau$ (i.e.\ SFR
$\propto \exp(-t/\tau)$),
% and SFR=constant with $\tau=\infty$),
the age $t_\star$ defined since the onset of star-formation, and
attenuation $A_V$ described by the Calzetti law
\citep{calzettietal2000}.
In addition, we here introduce a variable \lya\ strength described
by the {\em relative} \lya\ escape fraction \frellya $\in [0,1]$, defined
by 
\begin{equation}
L(\lya) = \frellya \times L^B(\lya), 
\end{equation}
where $L^B(\lya)$ is the intrinsic \lya\ luminosity of the spectral
template given by its Lyman continuum flux and the case B assumption,
and $L(\lya)$ is the adopted \lya\ luminosity for the spectral template
(before any additional attenuation with the Calzetti law, assumed to 
affect stars and gas in the same manner).
Values $\frellya < 1$ therefore describe an {\em additional} reduction
of \lya\ beyond the attenuation suffered by the UV continuum.  

In practice we compute SED fits for all combinations of
%redshift 
$z \in $ [0,10] in steps of 0.1, 
%metallicities 
$Z=$ (0.02=\zsun, 0.004, 0.001),
$\tau=$ (10, 30, 50, 70, 100, 300, 500, 700, 1000, 3000, $\infty$) Myr,
51 age steps from 0 to the age of the Universe \citep[see][]{bolzonellaetal2000},
$A_V \in $ [0,4] mag in steps of 0.1, and
$\frellya =$ (0,0.25,0.5,0.75,1.0).
Minimisation over the entire parameter space yields the best-fit parameters as well as
other properties such as the predicted \lya\ equivalent width and UV magnitude.
To determine the uncertainties of the physical parameters, we use Monte-Carlo simulations
by running typically 1000 realisations of each object. From this we derive the probability 
distribution function for each parameter/quantity, either for each individual object or for (sub)samples.

Examples of best-fit SEDs of two B-drop galaxies where a high (low) \lya\ equivalent
width is inferred from the data are shown in Figs.\ \ref{fig_sed1} (\ref{fig_sed2}) for illustration.
The position and shape of the  availabe photometric bands are also shown.
Several issues can be seen from these figures.
First, we see that 2-3 bands, unaffected by \lya, are available to constrain
the restframe UV domain, i.e.\ the UV slope. This is also true for our higher redshift
samples, where e.g.\ the JHK bands probe the UV restframe for I-drops. Of course,
the photometry is not necessary available and deep enough for all objects. 
Second, in general the contribution of nebular emission to the broad-band filters can vary strongly 
from case to case. See Fig.\ \ref{fig_sed1} showing an object best fit with a strong 
lines and a strong nebular continuum, whereas for the object shown in Fig.\ \ref{fig_sed2} 
nebular emission is negligible. Note, that the first object shows a clear excess at
3.6 \micron\ with respect to 4.5 \micron, which is naturally explained by presence
of strong lines (\ha\ and others) in the first filter and few lines in the second
for galaxies in this redshift range \citep[$z \sim$ 3.8--5, cf.][]{shimetal11,dBetal11}. 
In this case strong intrinsic \lya\ emission is also expected, and our fitting method
yields a median (mean) \lya\ escape fraction \flya=0.75 (0.57) and a corresponding
\lya\ equivalent width \ewlya=93 (71)  \AA.
Finally, strong \lya\ emission does not necessarily imply a strong excess in the filter
encompassing this line with respect to the next redward filter, since intrinsically
\ewlya\ is not very large, and since the continuum flux starts to be reduced
by the IGM at $z\ga3$.

Typically \lya\ contributes to $\sim$ 25-30\% of the broad-band flux for U- to I-drops 
if we assume EW$_{\rm rest}(\lya)=60$ \AA\ and compare this value to the
filter width. However, this contribution is higher if we take into account the fact
that the continuum flux decreases blueward of \lya. Therefore, depending 
on the source redshift and on the exact IGM attenuation, the relative contribution
of \lya\ to the photometric signal can significantly exceed the above estimate.
In any case, when sufficient filters are available, the \lya\ signal can be 
determined (albeit with considerable uncertainty), as we will show below.

%%%%%%%%%%%%%%%%%%%%%%%%%%%%%%%%%%%%%%%%%%%%%%%%%%%%%%%%%%%%%%%%%%%%%%%%%%%%%%%%%
\section{Results}
\label{s_results}

\subsection{Evolution of the \lya\ fraction with redshift and UV magnitude}

% % % % % % % % % % % % % % % % % % % % % % % % % % % % % % % %
\begin{figure}[htb]
\centering
\includegraphics[width=8.8cm]{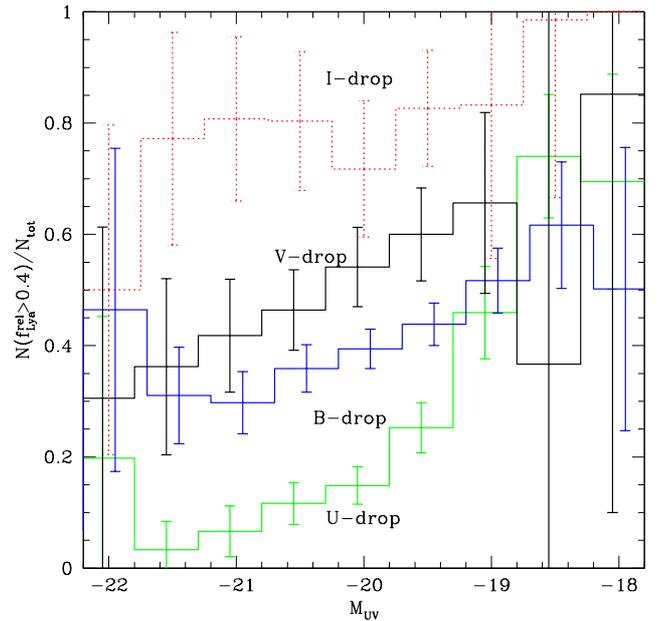}
\caption{Fraction $r_{\rm Lya}$ of galaxies with \lya\ emission (defined by $\frellya >0.4$)
as a function of the absolute UV magnitude (at 1500 \AA) derived
from our U (green line), B (blue), V (black) and I (black dotted) drop samples. }
\label{fig_lya_muv}
\end{figure}
\begin{figure}[htb]
\centering
\includegraphics[width=8.8cm]{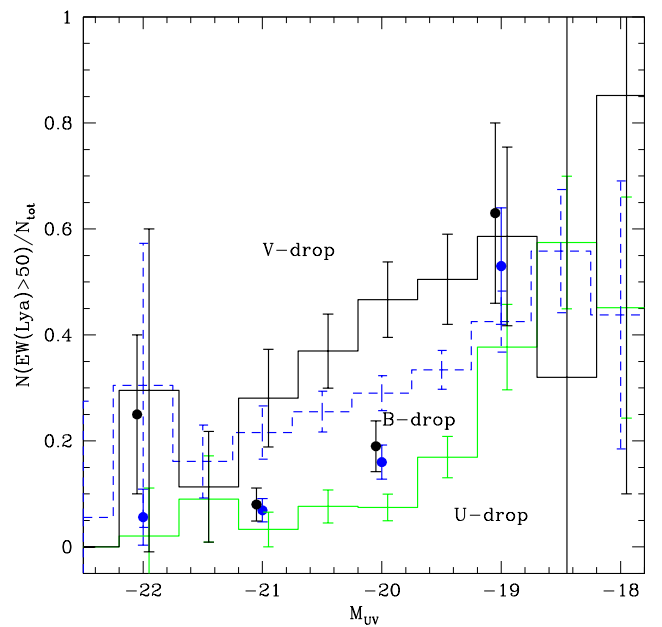}
\caption{Same as Fig.\ \protect\ref{fig_lya_muv} for galaxies with a large 
\lya\ equivalent width (\ewlya $>$ 50 \AA) compared to the observed fraction
derived from follow-up spectroscopy of $z \sim 4$ (blue points) and 5 
(black points) LBGs (data for B and V drops by from \citet{stark10}).
 }
\label{fig_lya2_muv}
\end{figure}
% % % % % % % % % % % % % % % % % % % % % % % % % % % % % % % %

We have examined the probability distribution function (pdf) of the 
 {\em relative} \lya\ escape fraction \frellya\ of our samples.
Overall it turns out that  \frellya\ is not well constrained for individual objects.
For each sample, however, the pdf shows two relative maxima close to  
$\frellya \approx 0$ and 1, whose relative importance varies between the samples.
To quantify this behaviour further, regardless of the detailed shape of the pdf, we count
the number of objects with a relative escape fraction above 40\%,
and define the corresponding fraction of \lya\ objects as $r_{\rm Lya} = N(\frellya >0.4)/N_{\rm tot}$.
This number should provide a simple estimate
of the fraction of objects with \lya\ emission among the total dropout samples.

In Fig.\ \ref{fig_lya_muv} we plot $r_{\rm Lya}$ as a function of the absolute UV magnitude
for the samples of $z\sim 3$, 4, 5, and 6 galaxies. 
%In practice, we have derived $r_{\rm Lya}$ from the full 2D pdf of $\frellya$ and $M_{\rm UV}$. 
Two main results emerge from this figure. 
First, we find that the fraction of objects showing \lya\ emission increases with redshift 
(both on average and within each magnitude bin). 
Second, at each redshift we find that \lya\ emission is more common in galaxies with fainter 
UV magnitudes.
Except at the brightest and faintest magnitudes, where the number of galaxies is relatively
small, our trends are significant. 
The same trend is also found when using cuts in equivalent width instead of $r_{\rm Lya}$, as shown below.

Interestingly our derived \lya\ fraction
shows the same behaviour with redshift
and UV magnitude as obtained from and hinted at by several other studies, mostly based on spectroscopic 
observations of $z\ga 3$ galaxies. 
For example, using photometric criteria derived from spectroscopy, \citet{cooke09} 
showed an increasing fraction of objects with strong \lya\ emission among faint
$z \sim 3$ LBGs.
The most direct comparison can be made with the work of \citet{stark10}, who determined 
the \lya\ fraction of LBGs from spectroscopy of $\sim$ 400 B, V, I, and z dropout galaxies. 
These authors found indeed the same trends just described both with UV magnitude and with
redshift. To compare our \lya\ with that derived by \citet{stark10} we apply the same criterium
on the \lya\ restframe equivalent width (\ewlya $>$ 50 \AA) on our samples.
The result, shown in Fig.\ \ref{fig_lya2_muv}, is very encouraging, although some discrepancy seems
to remain at intermediate magnitudes. The same holds when we compare our data with those
of  \citet{stark11}.
For our U-drop sample we find $\approx 10 \pm 5$ \% of objects with \ewlya $>20$ \AA\
at $M_{\rm UV} < -20.5$, comparable to $\sim$ 20--25 \% of LBGs in the sample of \citet{shapley03}
at similar UV magnitudes.
Among $z \sim 5$ LBGs \citet{douglas10} find $\sim$ 55\%  with \ewlya $>20$ \AA,
or $22 \pm 4$ \% after correction for spectroscopically unconfirmed galaxies.
Our values for the same \ewlya\ cut are between $\sim$ 15--60 \% for magnitudes
brigther than $M_{\rm UV} \sim -20$. 
We conclude that the \lya\ trends derived with our method 
%from broad-band SED fits
are in reasonable agreement with the available spectroscopic data.

Other data also supports the trend of a higher \lya\ fraction at high redshift. 
For example, a convergence of the luminosity functions and number densities
of LBGs and LAEs has been noticed by various studies \citep[e.g.][]{ouchi08,cassata11}.
Using such surveys \citet{hayes11,blanc10} have shown that, on average, the 
absolute \lya\ escape fraction\footnote{The absolute \lya\ escape fraction 
is given by \frellya $\times 10^{-0.4 A_{\rm Ly\alpha}}$, where $A_{\rm Ly\alpha}$ is the
attenuation at the wavelength of \lya.} increases with redshift from $z \sim$ 0 to 6.5
in UV selected, star-forming galaxies. Our analysis shows that this also holds for LBGs.

% % % % % % % % % % % % % % % % % % % % % % % % % % % % % % % %

\subsection{The distribution of \ewlya\ with magnitude and 
other comparisons}
The ``recovered'' best-fit values of \ewlya\ for all $\sim 1300$ galaxies from $z \sim$ 3--5 
are plotted  in Fig.\ \ref{fig_ewlya} as a function of the UV magnitude.
Our models show clearly -- at each redshift -- an absence of UV bright objects with 
large \lya\ equivalent widths, whereas for fainter objects \ewlya\
occupies a wide range.  This behaviour is well known, as it has been found
in virtually all samples of LAE and LBGs at $z \ga 2$ (or above 3)
\citep[see e.g.][]{shapley03,ando06,Tapken06,shimasaku06,ouchi08,pentericci09,stark10},
although \citet{nilsson09} have questioned the statistical robustness of this trend.
Our analysis of a large sample of LBGs therefore provides an independent confirmation 
for the deficit of large \lya\ equivalent widths at bright UV continuum luminosities 
and for an increasing maximum of \ewlya\ toward fainter magnitudes
found from spectroscopic and narrow-band surveys of LBGs and LAEs.
The statistical significance of our result has already been shown in Fig.\
\ref{fig_lya_muv}. 

% % % % % % % % % % % % % % % % % % % % % % % % % % % % % % % %
\begin{figure}[htb]
\centering
\includegraphics[width=8.8cm]{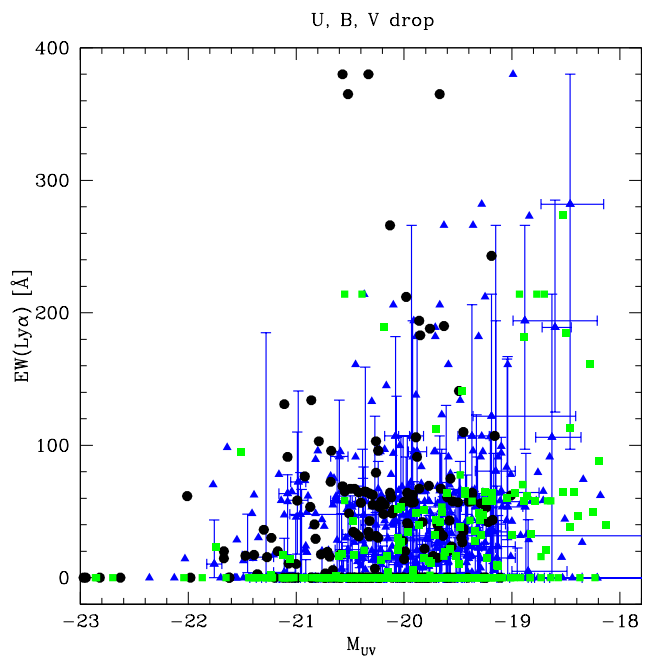}
\caption{Predicted best-fit values of the restframe \lya\ equivalent width \ewlya\ 
as a function of $M_{\rm UV}$ for the U, B, and V dropout galaxies studied here. 
Same colour symbols as in the previous Figures. To illustrate the typical 
uncertainties for the individual objects, we show errorbars randomly for every fifth
B-drop.}
\label{fig_ewlya}
\end{figure}

\begin{figure}[htb]
\centering
\includegraphics[width=8.8cm]{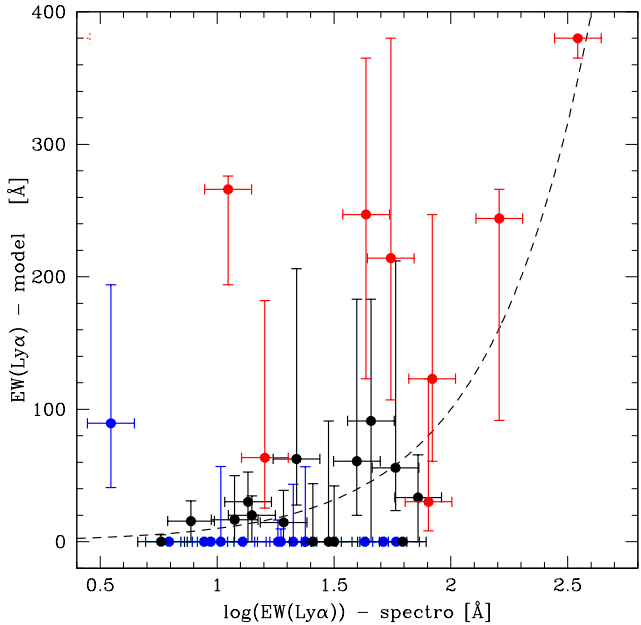}
\caption{Comparison of the predicted \lya\ (rest-frame) equivalent width with the observed one
for the 36 objects in common with spectroscopic measurements of \citet{stark10}. 
Observational errorbars of $\sim$ 25\%, typical for the brighter objects, are adopted.
A logarithmic x-axis is chosen to optimise the readability of the plot.
Red symbols indicate the i-drops, black V-drops, and blue B-drops.
For approximately 60\% of the objects we find an agreement within a 68\% confidence interval.
}
\label{fig_ewobs}
\end{figure}
% % % % % % % % % % % % % % % % % % % % % % % % % % % % % % % %

The \lya\ equivalent widths predicted from our SED fits span a reasonable range of values,
comparable to those observed in various LBG samples \citep[cf.\ e.g.][]{stark10,pentericci09,
shapley03}, with most values between 0 and $\sim$ 60 \AA\
(the typical value for constant star-formation with $\frellya \approx 1$).
A good agreement is found between the EW distribution of \citet{shapley03} and our
results for the U-drop sample.

Finally we have examined all individual objects for which spectroscopic measurements 
of \lya\ where available. From the data
of \citet{stark10} we find 36 objects in common with our sample. 
The predicted and observed \lya\
equivalent widths of these objects is shown in Fig.\ \ref{fig_ewobs}.
Overall we find an agreement within a 68\% confidence interval for $\sim$ 60\% 
of the objects. The main outliers are 3 i-drop galaxies for which our model
predicts too large equivalent widths (possibly related to uncertainties with the
IGM treatment, cf.\ below), and several B-drops where the model
underestimates \ewlya. Although encouraging, further comparisons with larger samples
may be helpful.

We conclude that our SED fitting method relying purely on broad-band photometry allows 
us to derive statistically meaningful trends of \lya\ with redshift and UV magnitude.
Furthermore the trends and the predicted \lya\ strength appears both qualitatively
and quantitatively in good/reasonable agreement with the existing observations.

\subsection{Test of our method using mock simulations}
The above comparison of our results with 36 galaxies with available \lya\
measurements can be improved by mock simulations allowing a more 
complete assessment of our method.

To do this we have constructed a set of synthetic galaxy spectra with
known physical properties including \ewlya,  from which we then
derive the synthetic photometry including typical observational errors.
The photometry of these mock galaxies is then fitted using our
SED fitting tool, and the derived \ewlya\ compared to the intrinsic (input) one.
In practice we have chosen to simulate a sample of 500 B-drop galaxies.
To construct their synthetic spectra we randomly draw their 
physical parameters, i.e.\ varying age from 0 to $\sim$ 0.5 Gyr, star-formation histories
with variable $\tau$ (taking the same values as for the SED fits, cf.\ above),
variable $\frellya \in [0,1]$, $A_V \in [0,1]$, and $z \in [3.8,4.5]$.
We chose the intermediate metallicity ($Z=0.004$), since our fits
are very degenerate in $Z$ \citep[cf.][]{dBetal11}.
Finally we add noise to the synthetic photometry
taking the median photometric uncertainty of the observed B-drop sample
in each band. Our artificial sample of galaxies should therefore be representative
of typical B-drop galaxies.

The mock sample generated in this manner is then fitted using  the same
procedure adopted for the observed LBGs, allowing us to verify the accuracy
of our fitting procedure. The result for the \lya\ equivalent width
is shown in Fig.\ \ref{fig_mock_ew}.  A good agreement
is found between the derived (output) and input equivalent widths
albeit with a relatively large uncertainty. 
Examing the full probability distribution function of the derived equivalent widths
we find (\ewlya$ -  EW_{\rm in}) = 0. ^{+45}_{-25}$ \AA\ (within 68\% confidence),
which demonstrates that our method is statistically able to recover the input value
with no systematic offset. As expected the dispersion between the input
and output equivalent widths increases with decreasing input \ewlya.
Note that the above dispersion has been obtained for a sample with a relatively
low median $EW_{\rm in}=20$ \AA. 

One may wonder whether the accuracy of recovering the strength of the
\lya\ line depends on UV slope of the galaxy. If this was the case the trends
found above with UV magnitude could be biased, since the observed
UV slope appears to correlate with this quantity 
\citep[e.g.][]{Bouwens09_beta}.
In Fig.\ \ref{fig_mock_ewdiff} we show the relative difference (``error'') between
the output (derived) and input \lya\ equivalent width for our mock sample
as a function of the UV slope $\beta$ determined in a standard fashion
from the (\iacs-\zacs) color \citep{Bouwens09_beta}.
For clarity we separate galaxies according to their input equivalent
width EW$_{\rm in}$ in this plot. We find no trend of the relative 
``error'' with $\beta$ for objects with EW$_{\rm in}>50$ \AA.
For objects with faint \lya\ emission the errors become larger, as expected,
but show no clear trend with UV slope, at least over the range 
of  typical slopes observed at $z \sim 4$ ($\beta \sim -3$ to $-1$). 
We therefore conclude that our multi-band SED fitting method
is able to recover the strength of \lya\ emission with an uncertainty
which is independent of the intrinsic UV slope of the galaxy.

% % % % % % % % % % % % % % % % % % % % % % % % % % % % % % % %
\begin{figure}[tb]
\centering
\includegraphics[width=8.8cm]{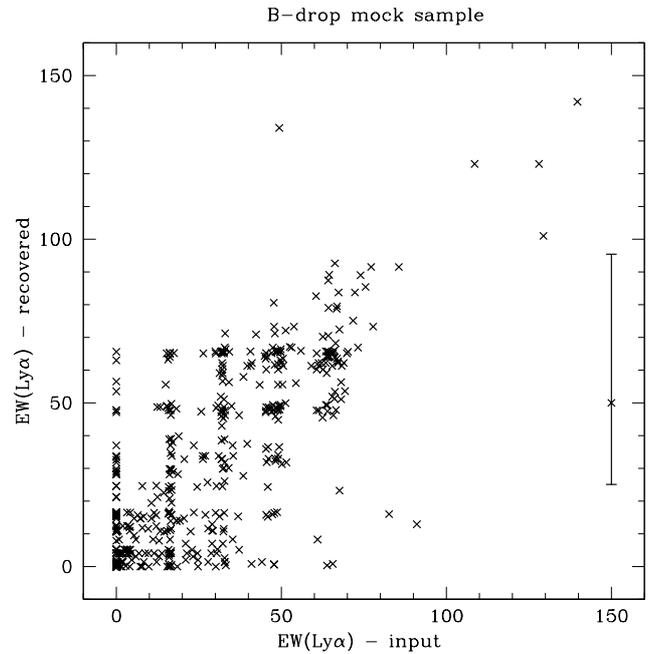}
\caption{Derived median \lya\ equivalent width (restframe) versus
input \lya\ equivalent width for our mock sample of 500 B-drop galaxies.
The errorbar indicates the median 68\% confidence limit on the derived
\ewlya; no offset between the derived median and the input value is obtained,
which demonstrates that our method works at least statistically.
A small fraction of objects is found at higher \ewlya, outside the plot.
A good agreement is obtained for all of them.
}\label{fig_mock_ew}
\end{figure}

\begin{figure}[tb]
\centering
\includegraphics[width=8.8cm]{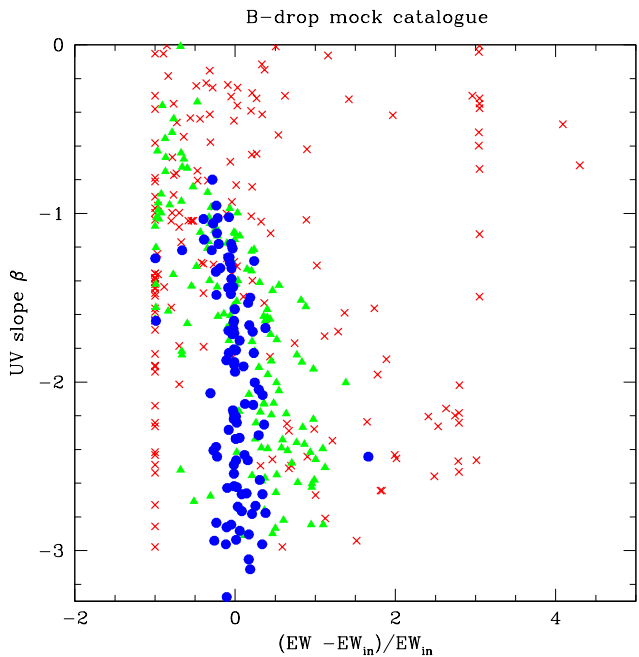}
\caption{Relative difference between
the output (derived) and input restframe \lya\ equivalent width for our mock sample
as a function of the UV slope $\beta$ determined from the (\iacs-\zacs) color.
Blue: objects with EW$_{\rm in} > 50$ \AA, green: $50 \ge {\rm EW}_{\rm in} > 20$, black:
EW$_{\rm in} < 20$ \AA.
}\label{fig_mock_ewdiff}
\end{figure}

% % % % % % % % % % % % % % % % % % % % % % % % % % % % % % % %
\subsection{Possible biases}
Could the main results obtained with our method, i.e.\ the increasing
fraction of \lya\ emission towards fainter UV magnitudes and the
increase of \lya\ with redshift among LBGs, be due to biases and not
intrinsic to the galaxy population?
We think that this is not the case for the following reasons.

First, our method is not biased by a possible trend of the UV slope
towards fainter magnitudes, at least for the objects with sufficiently
strong \lya\ emission (e.g. \ewlya$>$ 50 \AA).
Furthermore our photometric method yields the same behaviour
as spectroscopic measurements, which should be subject to
different biases, and which generally do not reach the same depths.
To avoid biases due to the difficulty of detecting faint \lya\ emission
from objects faint in the continuum, \citet{stark10} have measured
the fraction of LBGs at $z \sim$ 4--6 with \ewlya\ above an appropriate 
threshold, and we have applied the same cuts here (Fig.\ \ref{fig_lya2_muv}).
Recent, deeper observations of $z\sim$ 2--3 LBGs by N.\ Reddy (Private communication) 
also show a persistent increase of the \lya\ fraction towards faint objects once
corrected for this bias.

Finally, these trends could also be biased by the LBG selection, which can
be affected by the strength of \lya\ emission \citep[cf.][]{stanway07,stanway08,reddy08,reddy09}
As pointed out by \citet{stark10} this bias is minimised by simultaneously
targetting ``adjacent'' drop-out populations, e.g.\ U,B,V, and I-drops
as done here. In this way objects with strong \lya\ emission scattered out from
one selection window will be recovered by the selection of the adjacent drop-out.
Therefore  biases related to \lya\ affecting the colors used for the 
LBG selection should only influence the approximate redshift distribution of galaxies, 
but should not bias the overall sample with respect to the strength of \lya.

We conclude that the derived trends of \lya\ with magnitude and redshift
must be an intrinsic property of UV selected star-forming galaxies at
high redshift.

\subsection{Discussion}
What explains the observed trends of \lya\ with UV magnitude and redshift?
Based on radiation transfer models and on comparisons with
observations \citet{verhamme08,schaerer08} have shown that absorption by dust is 
the main process responsible for the observed diversity of \lya\ line profiles
and strengths among LBGs and LAEs. From this they suggested that variations of 
the dust content with galaxy mass, UV magnitude, and redshift should explain
the increasing fraction of \lya\ emission both with $z$ and with decreasing
brightness of the objects. Findings of a correlation between extinction and
the \lya\ escape fraction \citep[cf.][]{atek09b,hayes10b}  and the observed decrease
of the average UV attenuation with increasing redshift further support this
picture, even quantitatively \citet{hayes11}.
To examine this further, we have checked whether our SED models reveal any 
correlation between \frellya\ and $A_V$
%dust attenuation 
(or other physical properties). 
Beyond the one shown above with $M_{\rm UV}$, we have not found clear
additional correlations, the uncertainties/degeneracies being too large.
We can speculate that future studies of large galaxy samples with additional 
observational constraints, e.g.\ from deep surveys with large sets of filters, 
should be able to detect such  correlations.

Can our method be used to constrain cosmic reionisation? In principle yes,
since as shown above, we are able to determine the strength of \lya\
in a similar way as spectroscopic observations, at least for a large sample
of galaxies. From this we are then in principle able to determine the
fraction of \lya\ emitters, the \lya\ luminosity function, or other related
quantities, which can in turn be used to probe the evolution of the neutral
hydrogen fraction with redshift \citep[see e.g.][]{malhotra04,dijkstra07}.
However, we should be aware of the following possible limitations.
At high-$z$ ($z \gg 5$) the IGM significantly alters the SED shortward of 
\lya. In order to correctly predict the \lya\ strength from broad-band
SED fits one therefore also needs the most accurate treatment of the
\lya\ forest (in principle of stochastic nature) possible. At the same
time  there are some degeneracies between \lya\ strength, IGM, and also
redshift, which can all modify the flux in the broad-band filter encompassing \lya. 
It is therefore clear that our method should become less accurate at very high redshift.
Future applications to larger samples of high-$z$ galaxies observed in as many 
photometric bands as possible, and careful examination of the above mentioned
degeneracies will show whether our SED fitting method can provide useful
constraints also on \lya\ during the epoch of reionisation.

%%%%%%%%%%%%%%%%%%%%%%%%%%%%%%%%%%%%%%%%%%%%%%%%%%%%%%%%%%%%%%%%%%%%%%%%%%%%%%%%%
\section{Summary and conclusions}
\label{s_conclude}

Using an updated version of the \hyperz\ photometric redshift code of \cite{bolzonellaetal2000}
adding nebular emission to the spectral templates 
\citep[see][]{schaerer&debarros2009,schaerer&debarros2010}, we have analysed 
a large sample of Lyman-break selected galaxies at $z \sim$ 3--6 in the GOODS-S field,
for which deep multi-band photometry from the U-band to 8 \micron\ is available.
To our extensive exploration of the parameter space covering redshift, metallicity, star-formation
history, age, and attenuation whose results are discussed in detail in \citet{dBetal11},
we have added a variable \lya\ strength, described by a  {\em relative} \lya\ escape fraction \frellya.
We show that significant trends of \lya\ strength with redshift
and with UV magnitude can be inferred from broad-band photometric observations of 
large samples of galaxies using our method.
The validity of our method is also demonstrated by our simulations of mock galaxies.
Our method relies on fitting all the photometric bands  constraining the UV and
optical rest-frame SED. However, even with just 3 bands some \lya\ trends may
be detectable, as e.g.\ shown already by \citet{cooke09} for $z \sim 3$ galaxies.

We find an increase of the fraction of LBGs showing \lya\ emission both with redshift,
and towards fainter UV-restframe magnitudes at each redshift. Furthermore, we also infer
a decrease of the average and the maximum \lya\ equivalent width toward bright objects.
Our results are fully compatible with the trends found from spectroscopic observations
of individual LBGs and LAEs \citep[e.g.][]{ando06,ouchi08,stark10}, and with other more 
global trends of  the \lya\ escape fraction with redshift \citep{hayes11}.

In principle our method should also be applicable to derive the \lya\ properties of $z>6$ galaxies,
of particular interest for studies of the history of cosmic reionisation, and  in others contexts.
At least we have shown that even broad-band photometric observations can reveal some
information about \lya\ emission in high redshift galaxies.

%%%%%%%%%%%%%%%%%%%%%%%%%%%%%%%%%%%%%%%%%%%%%%%%%%%%%%%%%%%%%%%%%%%%%%%%%%%%%%%%%
\begin{acknowledgements}
We have appreciated stimulating discussions  with Matthew Hayes, Naveen Reddy, and other colleagues during
the spring-summer of 2011.
The work of DS and SdB is supported by the Swiss National Science Foundation.
DPS is supported by an STFC postdoctoral research fellowship. 

\end{acknowledgements}

\bibliographystyle{aa}
\bibliography{references}

\end{document}